%% file: main.tex
\documentclass{vgtc}                          

\ifpdf
  \pdfoutput=1\relax                   
  \usepackage{graphicx}                
  \DeclareGraphicsExtensions{.pdf,.png,.jpg,.jpeg} 
\else
  \usepackage{graphicx}                
  \DeclareGraphicsExtensions{.eps}     
\fi%

\graphicspath{{figures/}{pictures/}{images/}{./}} 

\usepackage{microtype}                 
\PassOptionsToPackage{warn}{textcomp}  
\usepackage{textcomp}                  
\usepackage{mathptmx}                  
\usepackage{algorithm}                 
\usepackage{algpseudocode}             
\usepackage{times}                     
\usepackage{cite}                      
\usepackage{tabu}                      
\usepackage{booktabs}                  
\usepackage{soul}

\newenvironment{tight_itemize}{\begin{itemize} \itemsep
-1.5pt}{\end{itemize}}

\onlineid{0}

\vgtccategory{Research}

\vgtcinsertpkg

\title{GeoSneakPique: Visual Autocompletion for Geospatial Queries}

\author{Vidya Setlur\thanks{e-mail: vsetlur@tableau.com} %
\and Sarah Battersby\thanks{e-mail: sbattersby@tableau.com} %
\and Tracy Wong\thanks{e-mail: tracywong@tableau.com}}
\affiliation{\scriptsize Tableau Software}

\teaser{
  \centering
  \includegraphics[width=\linewidth]{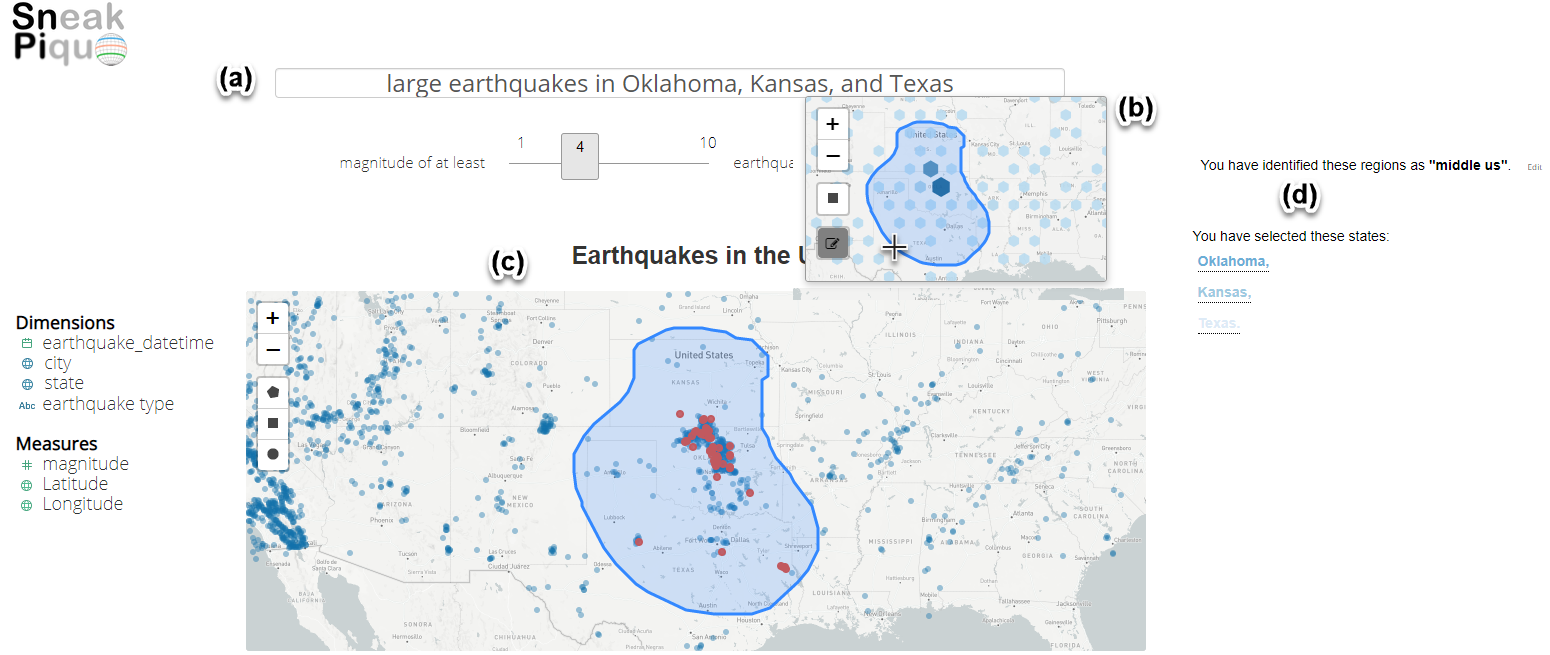}
  \caption{Screenshot of \textit{GeoSneakPique} showing a query for largest magnitude earthquakes (a) within a user-specified free draw polygon for a geographic area of interest (c).  The GeoSneakPique widget (b) provides a hexbin-based preview of the data distribution as well as a detailed basemap for additional spatial context.  The system interface provides feedback (d) on administrative geographic regions included (i.e., states) and allows for naming and saving the region for future queries ('middle us').}
  \label{fig:teaser}
}

\abstract{
How many crimes occurred in the city center?  And exactly which part of town is the `city center'?  While location is at the heart of many data questions, geographic location can be difficult to specify in natural language (NL) queries.  This is especially true when working with fuzzy cognitive regions or regions that may be defined based on data distributions instead of absolute administrative location (e.g., state, country). GeoSneakPique presents a novel method for using a mapping widget to support the NL query process, allowing users to specify location via direct manipulation \textit{with} data-driven guidance on spatial distributions to help select the area of interest.  Users receive feedback to help them evaluate and refine their spatial selection interactively and can save spatial definitions for re-use in subsequent queries.  We conduct a qualitative evaluation of the GeoSneakPique that indicates the usefulness of the interface as well as opportunities for better supporting geospatial workflows in visual analysis tasks employing cognitive regions. 
} 

\keywords{Data-driven scaffolds, cognitive region.} 

\CCScatlist{
  \CCScatTwelve{Human-centered computing}{Visualization}
}

\begin{document}


\firstsection{Introduction}
\maketitle
\input{sections/0_intro}

\section{Related Work}
\input{sections/1_relatedwork}

\section{System}
\input{sections/2_system}

\section{Evaluation}
\input{sections/3_evaluation}

\section{Discussion and Future Work}
\input{sections/4_discussion}

\section{Conclusion}
\input{sections/5_conclusion}

\newpage
\bibliographystyle{abbrv-doi}

\bibliography{bibtex}
\end{document}

%% file: sections/0_intro.tex
Information-seeking referring to the notion of \emph{place} is a prevalent form of human enquiry~\cite{Herskovits1997,landau_jackendoff_1993}. Despite the ubiquity of place in information-seeking, the \emph{semantics of place} is often subjective as the interpretation varies among people and how they relate to place~\cite{Montello2003}. As users formulate information goals, they often translate vague conceptual knowledge into more concrete descriptions. This translation can be difficult, however, as the `concrete' description from a user may not match the structure or content of the underlying data. When user input does not match or cannot easily be put into words as an NL query, the search process is often unsatisfactory. Search interfaces can help with some of the challenges related to this exploratory form of sense-making through user interface scaffolds such as autocompletion~\cite{belkin:1986}. Autocompletion displays in-situ suggestions as users input queries in the flow of their search tasks. These suggestions provide feedback to the user aiding them in generating valid queries with visual cues based on the underlying document corpora.  

In visual analysis, place is a basic category often employed to individuate meaningful portions of spatial locations during data exploration~\cite{maceachren2004maps}. With the proliferation of NL tools for visual analysis~\cite{thoughtspot,askdata,powerbi,setlur2016eviza,gao2015datatone,srinivasanorko,dhamdhere2017,sun2010articulate,hoque2017applying,yu2019flowsense}, users can express their analytical questions in plain language containing attributes and values from the underlying data source. Similar to web search tools, visual analysis NL interfaces also provide autocompletion to help users formulate queries~\cite{Marchionini:2006}. While these systems can enhance a user's ability to more easily, and successfully, generate NL spatial queries about specific, named locations (e.g., states, provinces, countries), there are still many areas ripe for exploration to allow users a more natural and flexible mode of spatial exploration that better aligns with the vague ways in which people often conceptualize space. 

Spatial language is complicated and there are numerous issues with identifying the intended meaning of spatial prepositions and relationships in NL usages~\cite{Talmy1983, Tversky1998}. The vagueness and ambiguity of expressing place-related terminology is commonly due to two considerations \cite{semple1907geographical,minshull1967regional,martin2005all}: First, generic place terms such as `area' and `region' are typically ambiguous in that their meaning is compounded from a number of distinct, but closely related senses. Second, concepts of place are often dependent on other concepts, such as geographic feature types, which are vague themselves.

 In our work, we explore how vague definitions for places can be expressed in visual autocompletion widgets through the concept of more concrete specifications of \emph{cognitive region}~\cite{Montello2003}. Cognitive regions are (approximately) two-dimensional features that people use to understand the (near) earth surface, as well as to reason and communicate about it.  These regions are spatial categories that often correspond non-arbitrarily to real entities, properties, and processes, and are created as intellectual or cognitive actions. They are a useful form of regionalization that correspond more readily to the reality of a heterogeneous set of geospatial features surface or serve the needs of a particular geospatial inquiry (e.g., ‘The Midwest’, `West Coast', and ‘downtown’). They may have irregular boundaries or may align nicely with common administrative boundaries (e.g., one definition of ‘West Coast’ may encompass all of Washington, Oregon, and California, while another definition of the same named region may simply be the land along the coastline). They may also be identified on-the-fly based on perception of data distributions (e.g., an interesting cluster of data points that are grouped into an arbitrarily shaped area of interest). Cognitive regions are particularly well-suited for NL interaction of geospatial data as they reflect the type of categorical thinking that so highly characterizes human thought and communication.  They also may be formed, or re-shaped, on-the-fly through evaluation of data distributions, with the boundaries of a region of interest expanding or contracting based on how data is distributed around the user's initial conceptual boundary.

\subsection{Contributions}
This paper introduces \textit{GeoSneakPique}, a system that supports the querying of named regions as well as arbitrary combinations of geographic regions, cognitive regions, or data-driven regions that cannot easily be represented in NL. Our contributions are as follows:
\begin{tight_itemize}
\item GeoSneakPique provides an aggregated `sneak peek' (hex bins, dual encoded to reflect count using color and size) of the data with a map widget to facilitate queries based on data distribution or using contextual spatial information from the detailed basemap.
\item We introduce a `coverage' metric to help users assess and refine spatial queries using commonly named administrative geographies as well as data characteristics. 
The system also persists spatial definition of named regions for use in future queries. This ensures consistency in analytics and facilitates comparisons between regions.
\item  An evaluation of the system provides useful insights for future system design of NL input systems for supporting geospatial inquiry involving cognitive regions.
\end{tight_itemize}

%% file: sections/1_relatedwork.tex
\subsection{Cognitive regions}
 While the concept of location is fundamental in geography and facilitates our categorization of locations and attributes, it can be tricky to make a clear match between the human understanding of a location and a computer mapping of the location \cite{tuan1977, couclelis1992}.  Incorporating cognitive regions or other locations with fuzzy or irregular definitions, is a known difficulty and important challenge in NL interfaces \cite{purves2018}.  Montello \cite{Montello2003} suggests four distinct types of regions: administrative, thematic, functional, and cognitive.  These geographic regions may have sharp, well-defined, and official boundaries (e.g., states and countries), or vague and more personally relevant, conceptual definitions (e.g., `downtown’ or `west coast’), or they may be a combination of both (e.g., a neighborhood, which may have an official boundary defined by the city or county, but have a fuzzier border for individuals based on their personal categorization of location).  The regions are often fuzzy and vague, with substantial variation between individuals - even for the same named region (e.g., the boundaries of Northern and Southern California; \cite{montello2014vague}).  Additionally, another challenge in working with cognitive regions is that the precise definition of a named region may vary based on the way in which it is used or interacted with.  The boundary of the `west coast’ may have different meanings depending on the nature of the question being asked about the region - the region defined when asking about best surf breaks and the region used when asking about trends in agricultural production across the west coast will likely be different even though the named region (`west coast') is the same.

\subsection{Geospatial queries and expressing spatial concepts}
Map reading tasks typically fall into three categories - identifying specific information about locations, assessing general information about patterns across an entire region, or to facilitate comparisons between multiple locations or attributes \cite{slocum2009}. However, asking questions about location requires that we clearly define the location in question - for instance, a defined geography or a geographic name that can be attached to a known location (e.g., the term `California’ can be matched to a polygon with a name attribute of `California’). In writing spatial NL queries, it can be challenging to align a user’s name for a location to an absolute geographic definition.  This is a classic problem for NL queries as seen in toponym disambiguation research \cite{buscaldi2011}, as well as more broadly in understanding cognitive regionalization \cite{montello2014vague}. To further the challenges of specifying user locations in NL queries, the location of interest may not even have a common name and may be data driven, for instance, `the area around that cluster of data points over there' or `that land area sticking out near the lake'. Sketching has long been thought of as a natural way to express spatial information (e.g., \cite{egenhofer1997}) and has been incorporated into various systems as a support for defining location (e.g., graphical selection in Google maps~\cite{googlemaps}), spatial relationships (e.g., \cite{egenhofer1997}), or to query for specific geographic patterns / configurations (e.g., \cite{blaser2000visual, tang2020extracting}). 

\begin{figure*}[ht]
\centering
  \includegraphics[width=\textwidth]{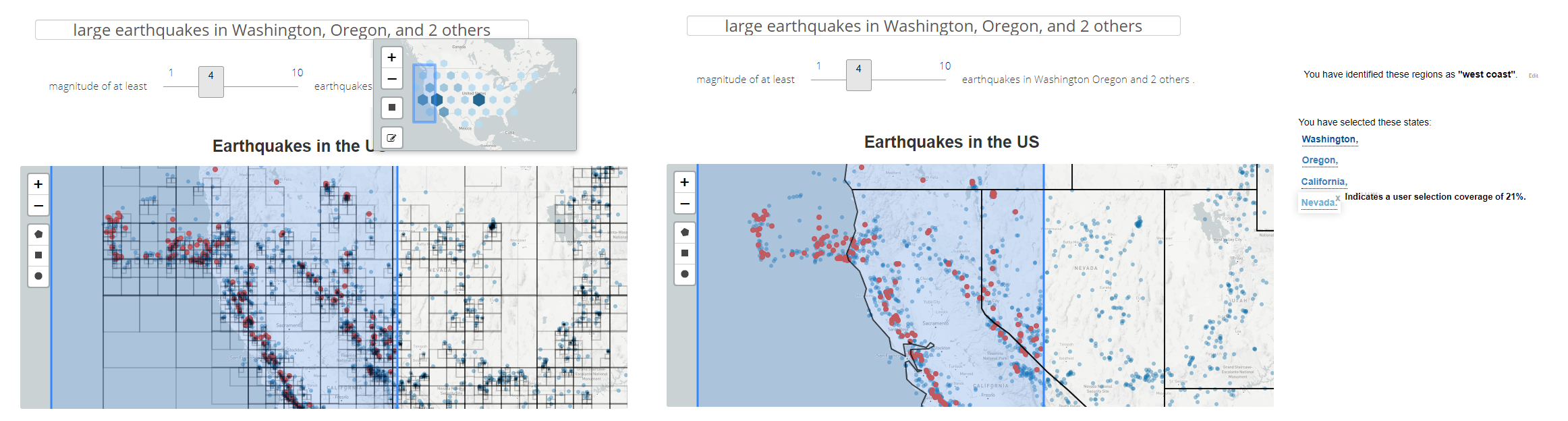}
  \caption{Left: Corresponding data points in the quadtree representation from the user's rectangular selection in the map widget. Right: Determine the coverage of state areas based on the user selected region.}
  \label{fig:quadtree_usergeom}
\end{figure*}

\subsection{Autocompletion and NL interaction}
Search and NL interfaces often employ text or visual autocompletion to help users formulate input queries~\cite{setlur:IUI2019,li:2009,yi:2017}. The autocompletion suggestions are either displayed contextually as a user types~\cite{grabski:2004,Qvarfordt:2013} or the interface reformulates the query into corresponding canonical expressions that represent the system’s language~\cite{setlur:IUI2019,askdata,powerbi,thoughtspot}. These scaffolds are useful in guiding the user to type syntactically complete and analytically valid queries during data exploration. However, these systems do not provide any preview of the underlying data, resulting in users having to determine questions of analytical interest, while formulating these questions in NL form. `Scented widgets’ demonstrated how some graphical user interface controls can support data analysis tasks~\cite{Willett07scentedwidgets:}. Their system enhanced traditional widgets such as sliders, combo boxes, and radio buttons with embedded visualizations to facilitate sense-making in information spaces. 

More recently, Sneak Pique~\cite{sneakpique} examined how both textual and visual variants of autocompletion with data previews provide users guidance within the context of NL interaction for visual analysis tasks. While Sneak Pique supports numerical, temporal, and spatial previews of the data, there are additional technical and linguistic challenges specific to supporting a fuller range of a user's spatial NL query needs. For instance, there 
is a classic geographic information retrieval problem in which the location(s) of interest in the user's query must be identifiable so they can be mapped to defined locations in the database~\cite{purves2018}. While Sneak Pique can enhance a user's ability to more easily and successfully generate NL spatial queries about specific, named locations, there are still opportunities to better support vague ways in which people often conceptualize locations. We extend the concept of data-driven scaffolds from Sneak Pique in \emph{GeoSneakPique}. We specifically explore how vague definitions for places can be expressed in visual autocompletion widgets through the concept of more concrete specifications of cognitive regions. 

%% file: sections/2_system.tex
\subsection{Overview}
GeoSneakPique employs a web-based architecture with the input NL query processed by an ANTLR parser~\cite{antlr} with a context-free grammar, similar to parsers described in~\cite{setlur2016eviza,hoque2017applying}. The parser accesses the dataset through the Data Manager to handle data query requests. Upon execution, the queries update the D3 Leaflet map~\cite{d3}. Similar to Sneak Pique~\cite{sneakpique}, the system polls the query as the user is typing and triggers grammar parse tree errors when the query is partially complete. Based on the underlying grammar rules, text- and widget-based auto completion suggestions are shown to the user to help resolve the partial queries. Given our specific focus on handling vague cognitive regions in the context of NL interaction, we extend the map widget to help users identify their region of interest in geospatial queries containing place-related tokens such as `near', `in', and `around'. The system also supports numerical and temporal descriptors in the queries such as `large', `small', and `recent'. The map widget provides a data preview and enables a user to select a region by either using a rectangular or free draw selection (Figure~\ref{fig:teaser}).

\subsection{Algorithm}
Algorithm~\ref{algo1} provides an overview of the algorithm for determining the coverage of the user selected cognitive region in the map widget.

\begin{algorithm}
	\caption{Determine coverage of selection}
	\hspace*{\algorithmicindent} \textbf{Input:} {Polygon object containing user selection}\\
	\hspace*{\algorithmicindent} \textbf{Output:} {List of geographies and their normalized scores}\\
	\hspace*{\algorithmicindent} \textit{qt} {is a quadtree data structure to store data points from dataset} \\
	\begin{algorithmic}[1]
		\State Visit \textit{qt} to get selected geo data points and the corresponding admin geography within the user selected region
		\State Get selected geo area and the corresponding admin geography unit (e.g. state) within the user selected region
		\State Take the aggregation of admin geography from selected geo data points and selected geo area
		\For {each admin geography in the aggregated list}
		    \State Calculate a normalized score given the proportion of geo data points selected and overlapping geo area (Eq.~\ref{eqn_score})
		    \If {score $<$ selected threshold}
		       \State Remove the admin geography from aggregated list 
		    \EndIf
	    \EndFor
	    \State Sort the scores of the aggregate list of admin geography in descending order

	\end{algorithmic}
	\label{algo1}
\end{algorithm}

\subsubsection{Compute normalized scores}
When a selection is made on the map widget, the algorithm uses the proportion of data points selected and the overlapping geographic area to determine the confidence level of selecting a particular geography. In our example, we use states, as county-level geography is too fine a unit and country-level too coarse.

To optimize for spatial queries, we use a quadtree, a compact data structure that facilitates search operations~\cite{klinger1971patterns}. We first perform a search on the quadtree to identify the selected points. For each state, we calculate the proportion of selected points to the total number of data points. We also calculate the proportion of geographic area for a state that intersects the user-defined region. Figure~\ref{fig:quadtree_usergeom} shows the intermediate results of the how GeoSneakPique calculates the proportion values. Lastly, we use both selected point proportion and overlapping geographic area proportion values to determine the confidence score. We adopted a heuristic approach and experimented with various individual weights for computing coverage of user selection. In practice, we found that assigning weights $0.65$ and $0.35$ to the overlapping geographic area and data points respectively, led to reasonable results to reflect likelihood of intentional inclusion of a specific geography. We found that a threshold of $0.2$ and higher worked well for choosing geographic areas that the user intended to include in their selection. Our observations experimenting with the various weights are documented in the supplementary material.
\hspace{-2mm}
\begin{equation} \label{eqn_score}
    \textrm{confidence score} = P_{area}*0.65 + P_{points}*0.35
\end{equation}

\subsection{User Interface}
Figure~\ref{fig:teaser} shows the GeoSneakPique interface with an input field for typing queries (a), a map widget for user selection (b), the main map view (c), and a panel to display the results of the targeted cognitive region (d). When a user selects a region in the map widget to complete a text query (e.g., ``large earthquakes in...''), the panel displays the various states sorted from the highest confidence score using a gradient color palette (Figure~\ref{fig:quadtree_usergeom} - Right). The user can choose to remove places that they do not want to associate with the selection as well as give the region a name in the text field provided.  The named region is saved by the system and can be referenced in future queries (e.g. ``what are the recent ones in the \emph{midwest}?''). The main map is updated to show the result from the query. 


GeoSneakPique also supports comparisons between two user-identified cognitive regions (e.g., ``compare \emph{the west} and \emph{the east}''). The system displays statistics minimum, maximum, and average values in each of these regions. The various system behaviors and query examples are demonstrated in the supplementary video.


%% file: sections/3_evaluation.tex
We conducted a user study of GeoSneakPique with the following goals: (1) collect qualitative feedback on how people express and query for cognitive regions in visual analysis and (2) identify system limitations and opportunities for how the semantics of place can be used to further data exploration. The study was exploratory in nature
where we observed the ways people explored data and how they responded to the system behavior. Because the main goal of our study was to gain qualitative insight in the system behavior, we encouraged participants to think aloud with the experimenter.

\subsection{Method}
\subsubsection{Participants}
We recruited $12$ volunteers (five males, seven females, age 36 – 65) from a local town mailing list. The participants had a variety of backgrounds  - user researcher, sales consultant, engineering leader, product manager, investor, commercial real estate broker, program manager, and marketing manager. Based on self-reporting by the participants, all were fluent in English and regularly used some type of NL search interface such as Google. Seven regularly used a visualization tool~\cite{tableau,powerbi} and the rest had limited proficiency.

\subsubsection{Procedure and Apparatus}
For the evaluation, we created a dataset of $\sim10,000$ earthquakes in the US ~\cite{usgsearthquake}, with a standardized structure and attributes\cite{usgsschema}. While we used earthquakes in our evaluation, the system will work with any point dataset. We began with a short introduction of how to use the system. Participants were instructed to phrase
their queries in whatever way that felt most natural and to tell us whenever the system did something unexpected. Although GeoSneakpique could handle other analytical queries, we asked participants to specifically focus on geospatial ones as we wanted to better understand how they would explore the data based on place. We discussed reactions to system behavior throughout the session and concluded with an interview. Each session took approximately 30 minutes.

\subsubsection{Analysis Approach}
We employed a mixed-methods approach involving qualitative and quantitative analysis, but considered the quantitative analysis as a complement to our qualitative findings. 

%% file: sections/4_discussion.tex
Overall, participants were positive about the system and identified many benefits. Given that we used a US earthquakes dataset for the study, most questions were centered around the intensity and recency of earthquakes occurring in various geographic areas. Several participants were impressed with the system's ability to understand their fuzzy geospatial queries (“It's neat that I am not bound by the constraints of the state boundaries when I want to dig deeper” [P9]). The participants appreciated the functionality for specifying and saving cognitive regions in their analysis (“It's convenient to not have to type all the states every time I want to reference the east coast” [P2]). The total number of queries that participants typed ranged from $8$ to $20$ ($\mu = 10.4$). The number of times the map widget was used to select a geographical region ranged from $5$ to $11$ ($\mu = 7.4$). Most of the times when participants interacted with the map widget, they named and saved a cognitive region; the number of times ranged from $6$ to $8$ ($\mu = 6.8$). Participants reused these saved cognitive regions $4$ to $8$ ($\mu = 5.9$) times in subsequent analytical questions in their user sessions. The most common cognitive regions that participants named were `the west' ($47\%$), `northwest' ($38\%$), `south' ($12\%$), and `midwest' ($3\%$). The most common analytical queries were related to `large' ($42\%$ of the interactions), `small' ($31\%$), and `compare' ($25\%$) earthquakes, with the remaining for `recent.'  Comments relevant to this behavior included, ``I want to see if there are actually large earthquakes around the ring of fire. It's convenient to be able to use `west' when I ask questions'' [P4], ``I am able to be specific by asking for `New York', but also more vague and just do a broader brush stroke on the New England area'' [P10], and ``I used cognitive regions as bookmarks to refer back to and I don't have to remember precisely what I selected in that little map'' [P7]. All participants interacted with the sliders and drop-down menus in the text response to understand the system behavior.

 The study also revealed several shortcomings and provides opportunities for supporting queries involving cognitive regions:

\noindent \textbf{Control over the spatial resolution}: In GeoSneakPique, the hexbins in the map widget adjust based on map zoom for providing some user control over spatial resolution. However, participants expressed interest in more control over the spatial resolution of the hexagons in the map widget used to discretize the data. For example, $P3$ stated, ``There seems to be more earthquake activity by the coastal regions on the west when compared to the central valley. I would have liked to be able to see more of that detail so I could fine tune my region to refer to Coastal California.'' Future work should consider providing more data-driven control, matching the scale of a user’s analysis to the scale of the data, or perhaps, including other spatial aggregation options, such as heatmaps.

\noindent \textbf{Comparisons between cognitive region features}: GeoSneakPique supports quantitative comparisons between cognitive regions by providing statistics such as mean, average, minimum, and maximum values. However, participants expected richer comparisons between features and the ability to specify which features they were interested in. $P11$ said, ``I am a commercial real-estate broker and have certain areas that I keep an eye on. I would like to see price differences between regions based on proximity to public transport, square footage, and urban density.'' Many of the analytical tasks involving cognitive regions tend to involve comparisons of complex properties~\cite{modifiers}. There is a need for supporting users with interaction techniques to specify the properties of interest and for visual analysis tools to provide richer summaries of such comparisons.

\noindent \textbf{Recommendations based on cognitive region properties}: Visualization recommendation systems are highly data-driven and rely on users' past behavior and preferences. Interfaces that support analytical inquiry with cognitive regions provide a motivating scenario for recommending other cognitive regions that may have similar data characteristics. $P1$ explained where such recommendations could be useful in his work - ``I develop medicine distribution and treatment logistics in developing countries. We need to look at the trend in cases, population, and number of treatment centers. It would be helpful if your tool could recommend new cognitive regions that my team has to look into based on what we have already focused on.''

%% file: sections/5_conclusion.tex
This paper presents a technique for providing graphical auto-completion to support querying cognitive regions of interest that cannot easily be represented in NL. We introduce a `coverage’ metric to determine the user's regions of interest through direct manipulation. GeoSneakPique allows for persisting the definitions of these cognitive regions where users can label, refine and incorporate them in future queries in the interface. An evaluation of the system indicates that participants found the system to be intuitive and appreciated the ability to specify vague geographic regions in their NL inquiry. Feedback from interacting with GeoSneakPique identifies opportunities for employing cognitive regions in richer geospatial data exploration. As Sigurd F. Olson~\cite{Olson1956TheSW} expresses the aesthetics of nature through the notion of \textit{place} - ``I see the mountain ranges of the West and the high, rolling ridges of the Appalachians. I picture the deserts of the Southwest and their brilliant panoramas of color, the impenetrable swamp lands of the South. They will always be there and their beauty may not change, but should their silences be broken, they will never be the same.''